\documentclass[]{raa}
\usepackage{amssymb,amsmath}            
\usepackage{graphicx,times}
\usepackage{natbib}
\usepackage{color}

\begin{document}

\title{Photometric studies of two W UMa type variables in the field of distant open cluster NGC~6866}


\volnopage{}
        \setcounter{page}{1}

        \author{Y. C. Joshi
        \inst{1}
        \and Rukmini J.
        \inst{2}
        \and S. Joshi
        \inst{1}
        }
   \institute     
{
$^1$Aryabhatta Research Institute of Observational Sciences, Manora Peak, Nainital, India - 263002 \\
$^2$Center for Advanced Study in Astronomy, Osmania University, India
}
%
%

%
\abstract{We present photometric analysis of the two W UMa type binaries identified in the field of distant open star cluster NGC~6866. Although these systems, namely ID487 and ID494, were reported in the Joshi et al. (2012), but a detailed study of these stars has not been carried out earlier. The orbital periods of these stars are found to be 0.415110$\pm$0.000001 day and 0.366709$\pm$0.000004 day, respectively. Based on the photometric and infrared colours, we find their respective spectral types as K0 and K3. The photometric light variations of both the stars show O'Connell effect which could be explained by employing a dark spot on the secondary components. The $V$ and $I$ bands light curves are analyzed using the Wilson-Devinney (WD) code and relations given by Gazeas (2009) which yield radii and mass of the binary components of the star ID487 as $R_1 = 1.24 \pm0.01 R_\odot$, $R_2 =  1.11 \pm0.02 R_\odot$, and $M_1 = 1.24\pm0.02 M_\odot$, $M_2 =  0.96 \pm0.05 M_\odot$ and for the star ID494 as $R_1 =  1.22\pm0.02 R_\odot$, $R_2 =  0.81\pm0.01$ $R_\odot$, and $M_1 =  1.20\pm0.06$ $M_\odot$, $M_2 =  0.47\pm0.01$  $M_\odot$.
}

 \authorrunning{Joshi et al. }            
 \titlerunning{Photometric studies of two W UMa stars in NGC~6866}  
 \maketitle

\section{Introduction}
\label{sec:intro}
The photometric and spectroscopic studies of eclipsing binaries (EBs) are important to determine their physical properties. These stars play an important role in testing stellar evolutionary models as they can provide mass and radius measurements for both the components of the binary system (e.g., Guinan et al. 2000, Torres \& Ribas 2002). These systems are also used to determine distances of the host star clusters (e.g., North et al. 2010, Kaluzny et al. 2013). One of the class of EBs is W Ursae Majoris (W UMa)-type stars whose light curves show continuous brightness variations, strongly curved maxima and minima with nearly equal depths. Binnendijk (1970) classified W UMa EBs into two sub-classes, A-type and W-type. The A-type binaries are of early spectral type from A to F and show deeper primary minima due to transit of larger, hotter component while W-type binaries are of late spectral type from G to K showing primary minima due to occultation of smaller, less massive component. In general, orbital periods of A-type stars are less than 0.3 day and that for the W-type stars, it typically lies between 0.3 day to less than a day. In the case of W-type stars, which are contact or over-contact binaries, both the components fill or over-fill their critical Roche lobes and are enclosed in a common convective envelope. The evolution of W UMa type binaries has been discussed with an ample detail in several studies in the past (e.g., Vilhu 1982, Eggen \& Iben 1989, Bradstreet \& Guinan 1994).

It is well understood that the fundamental physical parameters obtained for the W UMa binaries could aid and constrain the evolutionary and asteroseismic models. Most of the W UMa binary light curves show O'Connell effect, where a difference between the brightness at the maxima is observed despite having the similar effective temperatures for both the components of the binary system (O'Connell, 1951). This asymmetry is often associated with the presence of spots on one or both the components and is quite important in understanding the light variations of W UMa binaries. Such asymmetries have been found in many W UMa binaries e.g. BX Peg (Lee et al. 2004), CE Leo (Kang et al. 2004), FG Hya (Qian \& Yang 2005), QX And (Qian et al. 2007), FO Hydra (Prasad et al. 2013), among others.

The study of W UMa type binary stars in the star clusters is important in understanding their origin and evolution as well as to study the properties of the parent cluster. A number of contact binaries have been discovered in both galactic and globular clusters (e.g., Liu et al. 2007, 2008, 2011, Qian et al. 2006, 2007, 2015, Sriram et al. 2010, Zhu et al. 2014), however, only a few of them were investigated in a great detail. At ARIES, Nainital, we have been carrying out a long-term observational programme to search and characterize the variable stars in the field of open star clusters using various 1 to 2-m class telescopes in India (Joshi et al. 2012, 2014). The advantage of having such observations is that they cannot only be used to study the physical properties of the clusters but also to analyse some of the interesting variable stars lying in these clusters. In Joshi et al. (2012), we performed a photometric study of the intermediate-age open cluster, NGC~6866 (R.A. = 20:03:55, Dec. = +44:09:30). Interestingly, NGC~6866 is one of the four open clusters that lies in the field of Kepler survey and has been photometrically studied in a great length in recent times (e.g., Molenda-\.{Z}akowicz et al. 2009, Joshi et al. 2012, Balona et al. 2013, Janes et al. 2014, Bostanci et al. 2015). From the periodicity analysis of the photometric data taken over a period of 3 years, we detected 3 W UMa type binaries in the field of this cluster. Since no individual study has been carried out so far to derive the system parameters for these binary stars, therefore, we here carry out a detailed photometric analysis of two relatively brighter W UMa stars among them. The details of the data acquired for these EBs are given in Sec. 2 followed by a periodogram analysis in Sect. 3. The light curves analysis and estimation of their basic parameters are described in Sect. 4. The membership issue of these stars in the cluster is explored in the Sect. 5. Our results are discussed in Sect. 6.

\begin{figure*}
\centering
\vspace{-1.5cm}
\includegraphics[angle=0,width=16cm, height=14cm]{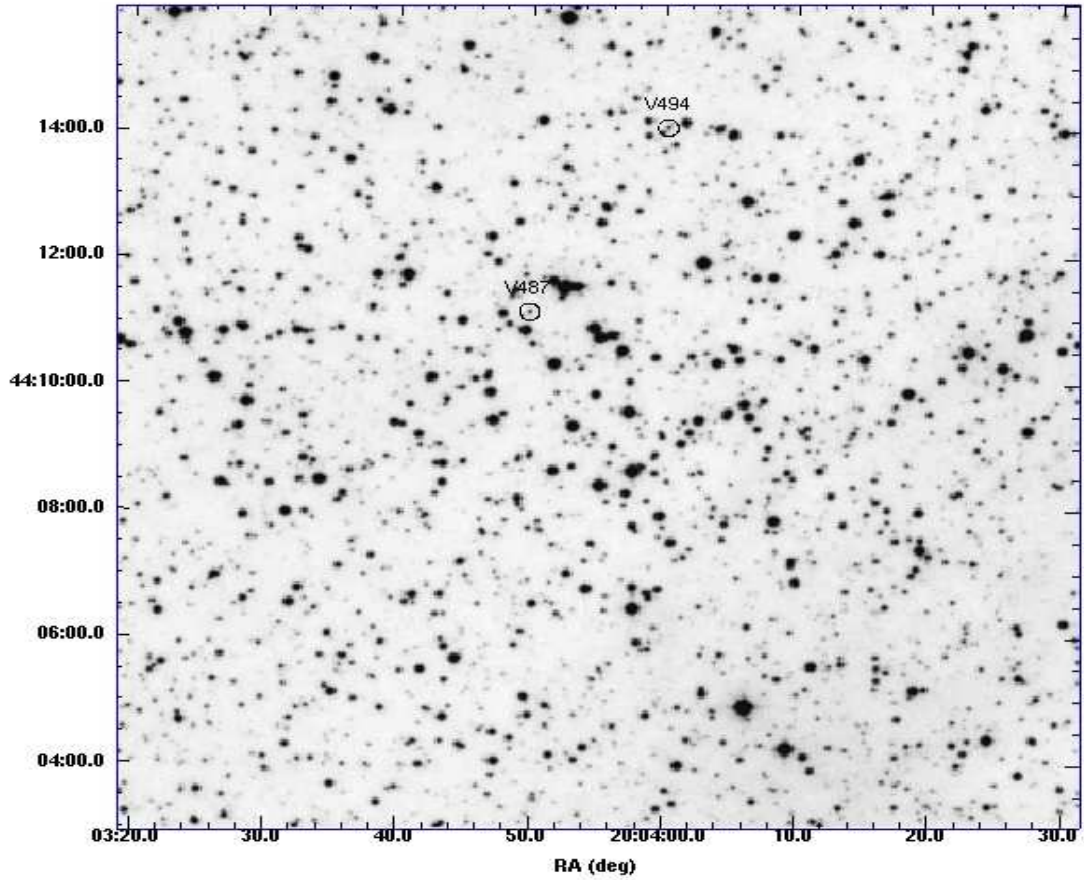}
\vspace{-1.3cm}
\begin{flushleft}
\caption{A $13' \times 13'$ finding chart for the two W UMa stars ID487 and ID494 in the field of cluster NGC~6866. North is up and east is to the right.}
\label{Fig1}
\end{flushleft}
\end{figure*}

%
\section{The Data}
The time-series photometric observations of stars in the field of intermediate age open cluster NGC~6866 were obtained  in Johnson $V$ and Cousin $I$ bands using 1.04-m Sampurnanand telescope, Manora Peak, Nainital for 29 nights between September, 2008 to January, 2011. The details of the observations and calibrations can be found in Joshi et al. (2012). From these observations, we detected 3 W UMa binaries, out of which two were bright enough to study further. The finding chart for these two stars is shown in Figure~\ref{Fig1}. Here, we also used the $V$ and $I$ band photometric data from the Molenda-\.{Z}akowicz et al. (2009). A total number of 713 and 45 data points were accumulated during our observations for the star ID487 in $V$ and $I$ bands, respectively, over a period of 831 days while a total of 434 and 458 data points were taken from  Molenda-\.{Z}akowicz et al. (2009) for these stars observed over a period of 85 days in the same passbands. For the star ID494 we acquired a total of 700 and 43 data points in $V$ and $I$ bands, respectively while 454 and 456 data points were taken from Molenda-\.{Z}akowicz et al. (2009) in the respective bands.
\section{Orbital period investigation}
\begin{figure*}[h]
\centering
\includegraphics[angle=0,width=14cm, height=10cm]{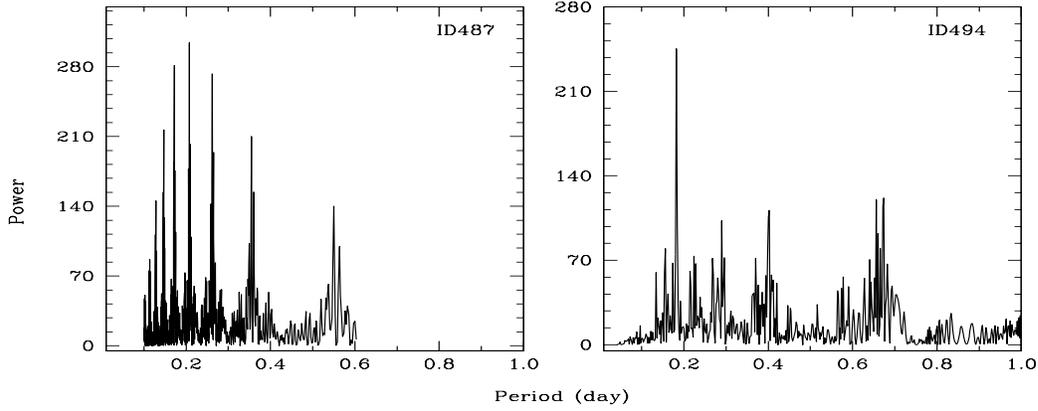}
\begin{flushleft}
\vspace{-4.5cm}
\caption{Power spectra obtained from the $V$ band data of the star ID487 (left panel) and ID494 (right panel).}
\label{Fig2}
\end{flushleft}
\end{figure*}

The estimation of precise orbital period is very important in the case of close binary system to yield information about the important physical processes such as evolutionary stage, mass transfer, magnetic activity cycles as well as to estimate accurate physical parameters of the binary system. Molenda-\.{Z}akowicz et al. (2009) first reported a period of 0.415010 day and 0.366528 day for the stars ID487 and ID494, respectively. In Joshi et al. (2012), we used a Lomb-Scargle method (Lomb 1976, Scargle 1982) to determine the period of all the variable stars in the unevenly spaced data which yield a period of 0.415110 day and 0.366704 day for the binary systems ID487 and ID494, respectively. In the present study, we further performed period analysis using Period04 (Lenz \& Breger 2005). The resulting power spectrums are shown in Figure~\ref{Fig2} where the peak periods are found to be at 0.207555 day and 0.1833545 day for the stars ID487 and ID494, respectively. Since EBs are better represented by two sine waves and therefore multiplying the period by 2 gives a true period of the binary system. We therefore assigned periods of 0.415110$\pm$0.000001 day and 0.366709$\pm$0.000004 day for the stars ID487 and ID494, respectively. Our present period estimates are quite in agreements with our earlier estimations in Joshi et al. (2012). We summarize the basic parameters of the two stars in Table 1.
%

 \begin{table}
    \caption{Details of the W UMa binaries taken from Joshi et al. (2012).}
    \begin{center}
\begin{tabular}{ccccccc}
\hline
W UMa& RA   & DEC   & $<V>$  &  $\Delta$V & (B-V) & Period   \\
ID   &       (J2000)      &  (J2000)            &  (mag) &  (mag)     & (mag) & (day)    \\ \hline
\\                                                                     
ID487                & 20:03:49.82 & +44:11:08.5  & 17.21  & 0.34       & 0.93  & 0.415110  \\
ID494                & 20:04:00.17 & +44:14:03.2  & 17.26  & 0.44       & 1.03  & 0.366709  \\
\hline
\end{tabular}
\end{center}
    \label{Tab1}
\end{table}

\begin{figure*}[h]
\centering
\includegraphics[angle=0,width=14.0cm, height=12.0cm]{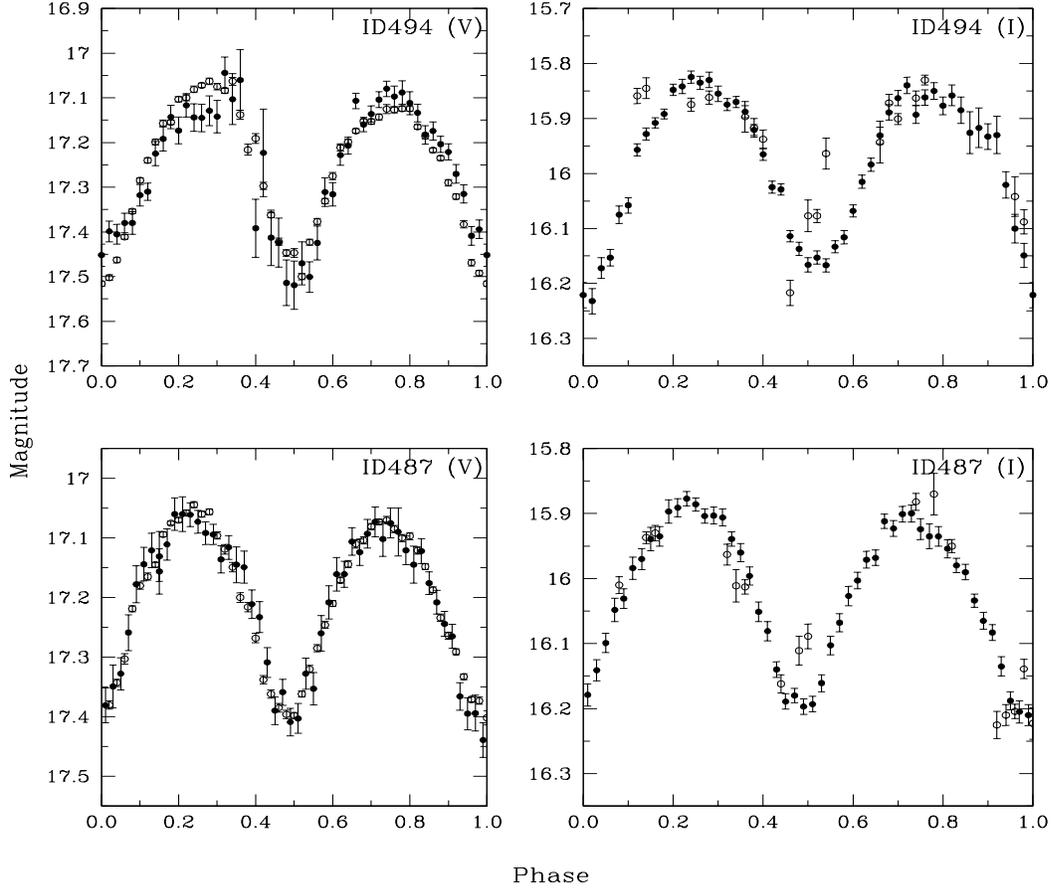}
\vspace{-0.3cm}
\begin{flushleft}
\caption{Folded $V$ and $I$ bands phase light curves for the W UMa stars ID487 (lower panel) and ID494 (upper panel). Open circles are data points obtained in the present study and filled circles are those taken from Molenda-\.{Z}akowicz et al. (2009) which are shifted to scale the present data.}
\label{Fig3}
\end{flushleft}
\end{figure*}

Using the periods given in Table~1, we derived the phase for each observations. Since there are large errors in the individual estimations, we binned the data in 50 bins of width 0.02 in phase. For those bins having photometric magnitudes and their respective photometric errors, the weighted mean magnitudes and errors were determined. Figure~\ref{Fig3} shows the folded light curves of stars ID487 and ID494 in $V$ and $I$ bands. The binned light curves have a smaller scatter, allowing a better visual identification of the binary light curves. The ephemeris for the minimum light of these binary stars can be represented by the following equations. \\

\noindent For the star ID487,
$$
Min~I (HJD) = 2455476.177535(55) + 0.415110(01) \times E
$$
For the star ID494,
$$
Min~I (HJD) = 2454736.283981(54) + 0.366709(04) \times E
$$
The initial epoch is one of our times of minimum light which was determined using the method given by Kwee \& Van Woerden (1956). The numbers in parentheses give the error in terms of last quoted digits.
%
\section{Photometric Analysis}

\subsection{Spectral Classification}
From the $UBVRI$ photometric study of the cluster field, Joshi et al. (2012) reported a reddened $(B-V)$ of 0.93 mag and 1.03 mag for the stars ID487 and ID494, respectively. Considering an interstellar extinction of $E(B-V)=0.10$ mag in the direction of the cluster NGC~6866 (Joshi et al. 2012), the unreddened $(B-V)_0$ was estimated to be 0.83 mag and 0.93 mag for the stars ID487 and ID494, respectively. We estimated the effective temperature of the primary component of the binary system using the following relation given by Wang (1994) for the contact binaries:
$$
logT_{eff} = 3.970 - 0.310~(B-V)_0
$$
This yields a temperature of 5160 K and 4805 K for the stars ID487 and ID494, respectively. Using the Table~15.7 given by Cox (2000), these temperatures correspond to the respective spectral class of K0 and K3 for the two W UMa binary stars.

From the 2MASS catalogue (Cutri et al. 2003), the infrared colour index $(J-K)$ for these two stars turns out to be 0.524 mag and 0.621 mag. The interstellar extinction is believed to be almost insignificant in the near infrared (Pribulla et al. 2009). Hence, assuming a negligible infrared reddening in the directions of the cluster NGC~6866 and taking the information from the Table~7.6 given by Cox (2000), we found an exactly similar spectral class of K0 and K3 for the two binaries. As latter spectral types are derived from the archival 2MASS photometry, and which are in resemblance with our observed photometric classifications, it can be concluded that our spectral classifications for these stars are quite robust.

\subsection{q-parameter estimation}
The mass ratio $(q = \frac{M_2}{M_1})$ of two components in a binary system is an important parameter to understand the evolution of a binary star. This is best determined from the spectroscopic analysis but our target stars are too faint ($\sim$ 17.2 mag) to have any spectroscopic data as yet. We therefore determined this parameter through the $q$-search method in the presently available photometric light curves using the WD code. To constrain $q$ parameter, the mean effective temperature of secondary component $T_2$, monochromatic luminosity of the primary component $L_1$, the dimensionless surface potentials of the two components $\Omega_1$ = $\Omega_2$, and the orbital inclination $i$ were adopted as adjustable parameters till a convergent solution was obtained for the binaries. To determine the exact value of $q$, this parameter was also taken as adjustable till a convergent solution was obtained.
\begin{figure*}[h]
\centering
\includegraphics[angle=0,width=14.0cm, height=10.0cm]{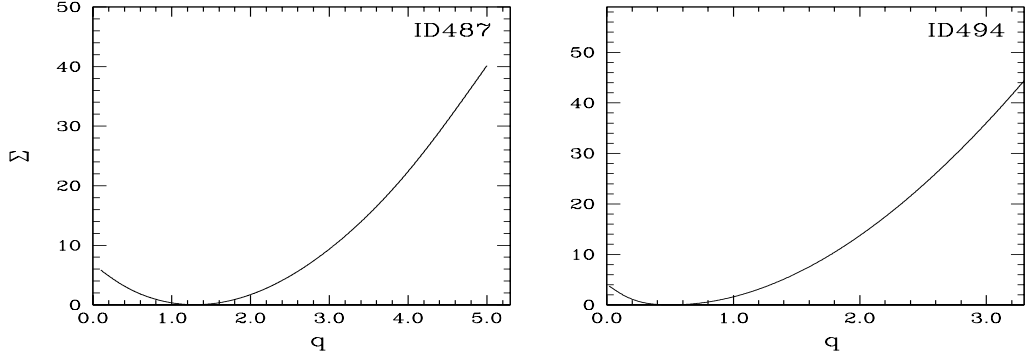}
\begin{flushleft}
\vspace{-5.5cm}
\caption{The plot of sum of squared residuals $\Sigma W(O-C)^2$ for different mass ratios. The left panel represents the variation for the binary star ID487 and right panel represents the variation for the star ID494.}
\label{Fig4}
\end{flushleft}
\end{figure*}
We determined sum of squares of the residuals $\Sigma W(O-C)^2$ for light obtained with different assumed values of $q$ starting from 0.1, in steps of 0.1. The corresponding values of potentials were calculated using the equations given by Kopal (1959). In Figure~\ref{Fig4}, we show the variation of $\Sigma W(O-C)^2$ as a function of $q$. An arbitrary smooth fit is drawn to connect the points and shown by a continuous line in the figure. The minimum value of $\Sigma$ in the plot represents the best value of mass ratio which is obtained as 1.3 and 0.5 for the stars ID487 and ID494, respectively.
		
\subsection{Light Curve modelling and Photometric solutions}
Light curve analysis was performed and photometric solutions were derived for the selected binaries using the Wilson and Devinney (WD) code v2003, which is a synthetic light curve analysis technique (Wilson and Devinney, 1971). In the WD code, some of the parameters need to be fixed and convergence of the solutions are obtained using the methods of multiple-subsets (Wilson \& Biermann 1976). As the contact configuration was evident in the photometric light curves for these two binaries, mode 3 of the WD code was found to be a better choice. Though, we initially started with mode 2 but the convergence was found to be more suitable for mode 3 in the light curve analysis.

Both the binaries showed the effective temperatures of later spectral types and the solutions indicate that they are W-subtype contact binaries with convective outer envelopes. Hence the gravity-darkening coefficients $G_1$ and $G_2$ for the components of the binaries were taken as 0.32 for the convective envelopes (Lucy 1967), the bolometric albedos $A_1$ and $A_2$ were taken as 0.5 for the convective envelopes (Ruci\'{n}ski 1969) and the limb-darkening coefficients $x_1$ and $x_2$ were taken as 0.6 and 0.8 for $V$ and $I$ bands, respectively (Al-Naimiy, 1978). The eccentricity $e$ of the orbit for the binaries was taken as 0, assuming circular orbits since the binaries are in contact configuration. Considering synchronous rotation and revolution of the W UMa binaries, we chose $F_1=1$. These parameters were kept fixed to obtain the optimal photometric solutions.

Luminosities were obtained from the observed light curves of the binaries in the respective bands using the amplitudes or depths of the minima, which were normalized to 4$\pi$. Considering the temperatures of the component, the model atmospheres were adopted for the binaries, hence we have chosen $IFAT_1 = 1$. The other parameters assumed were the third light $l_3 = 0$ and the longitude of periastron  $\omega = 90^0$ . In order to compute $L_2$ from $T_1$, $T_2$, $L_1$ and the radiation laws, the control integer $IPB$ was assigned as 0. 
 
\begin{figure*}
\centering
\includegraphics[angle=0,width=14.0cm, height=10.0cm]{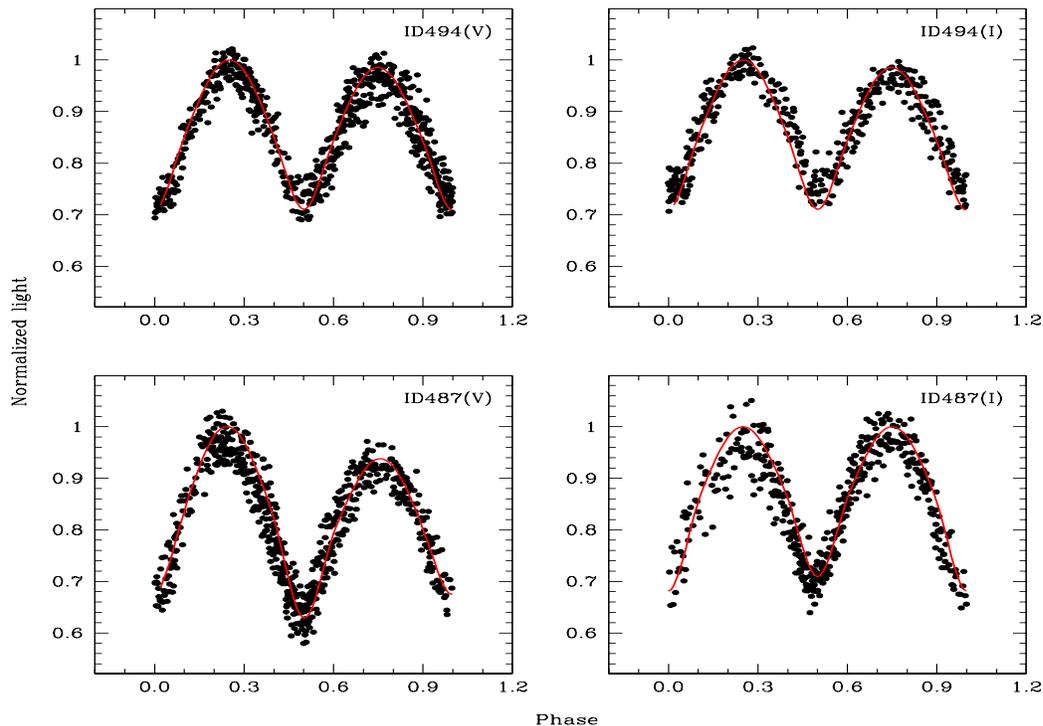}
\begin{flushleft}
\caption{Light curves of stars ID487 (top panel) and ID494 (bottom panel) in $V$ band (left side) and $I$ band (right side). The continuous lines represent the best fits obtained from the Wilson-Devinney solutions.}
\label{Fig5}
\end{flushleft}
\end{figure*}

The model fits were carried out as explained above using the DC program of the WD code following the procedure as explained in Rukmini et al. (2005) and Prasad et al. (2013). For each of the selected values of $q$ and the corresponding values of surface potentials $\Omega_1$, $\Omega_2$ by varying the adjustable parameters $i$, $L_1$, $T_2$ alternatively till the sum of the squared residuals $\Sigma W_i(O-C)_i^2$ showed a minimum and corrections to the parameters became smaller than their probable errors. The observed light curves show a clear O'Connell effect around the phase 0.75. Therefore, the spot models were adopted to explain the asymmetry of the light curves. After considering all possible spot models (hot or cool spots on the primary and secondary components), it finally turned out that adopting a cold spot on the secondary component of the binary systems resulted lowest residual in the solution of the WD program. We therefore took it as a best fit solution and results of our analysis are given in the columns 2 and 3 of the Table 2 for the $V$ band data of binary stars ID487 and ID494, respectively. Using the final parameters obtained from the solutions, the LC program of WD code was executed to obtain the theoretical light curves. The $V$ and $I$ bands light curves for the observed data points are shown in the Figure~\ref{Fig5} for the binaries ID487 (lower panel) and ID494 (upper panel). The fit of the theoretical curves along with that of the observed light curves are shown by continuous lines which seems to be satisfactory despite scatter in the data points due to faintness of the stars.

\begin{figure}
\centering
\includegraphics[angle=0,width=12.0cm, height=10.0cm]{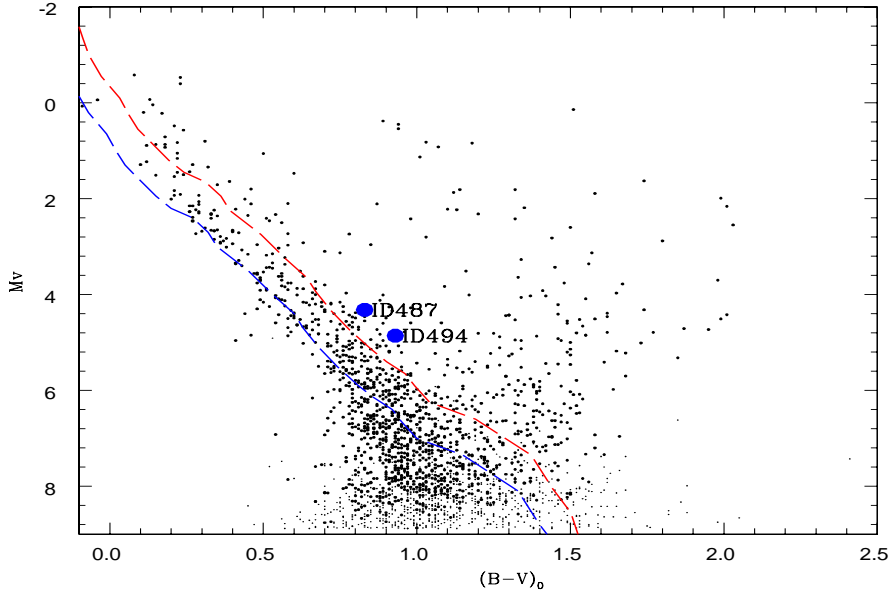}
\vspace{-2.0cm}
\caption{The $(B-V)_0$ vs $M_V$ colour-mag diagram for the stars in the cluster NGC~6866. The thick dashed line shown by blue line is the ZAMS corresponding to $V-M_V$ = 11.15  and E(B-V) = 0.10 mag, while the red dashed line indicates a shift in magnitude and colour due to binarity. Positions of W UMa binaries are shown by blue points in the CMD.}
\label{Fig6}
\end{figure}

\section{Distance estimation and cluster membership of W UMa stars}
W~UMa type variables are known to follow a PLC relationship between the period, color and absolute brightness (Ruci\'{n}ski 1969, Ruci\'{n}ski \& Duerbeck 1997) as given by:
$$M_V = -4.44 log(P) + 3.02(B-V)_0 + 0.12$$
\noindent
where $(B-V)_0$ is the intrinsic color and $P$ is the orbital period in days. As we have already determined $(B-V)_0$ and $P$ of the W UMa binaries, we derived their absolute magnitude $M_V$ as 4.32 mag and 4.86 mag, respectively for the stars ID487 and ID494.  Taking the visual magnitude of stars ID487 and ID494 as 17.21 mag and 17.26 mag, respectively, and assuming a uniform reddening of 0.10 mag in the directions of the cluster NGC~6866, we obtained a distance modulus of 12.58 mag and 12.09 mag for these stars in the respective order. This places them at a distance of 3.28 kpc and 2.62 kpc, respectively. In Figure~\ref{Fig6}, we draw the $(B-V)_0-M_V$ colour-magnitude diagram (CMD) where we show the positions of the stars identified in the field of the cluster NGC~6866. The intrinsic magnitude and color of the stars in the field of NGC~6866 were determined using the distance modulus $(m-M)_V$ = 11.15 mag and extinction $E(B-V)$ = 0.10 mag as estimated by Joshi et al. (2012). The dashed lines in the figure represent the ZAMS taken from Schmidt-Kaler (1982). We also over-plot the positions of two W UMa stars in the same colour-magnitude diagram with the parameters determined in the present study. It shows that the binary systems are above the main-sequence (MS) strip in the CMD of the cluster NGC~6866. According to Joshi et al. (2012), star ID487 has a very high proper motion probability of 0.85, spatial probability of 0.76 and photometric probability of 1.0 which makes it a likely-member of the cluster NGC~6866. However our analysis in the present study puts ID487 at the farther distance than the cluster NGC~6866 ($\sim$ 1.47) kpc and outside the MS strip suggesting that this W UMa binary may not belong to the cluster itself. The star ID494 has zero proper motion probability and a low spatial probability of 0.34 as given in the Joshi et al. (2012) which clearly suggests that it is an unlikely member of the cluster. The large distance of this star in comparison to that of the cluster NGC~6866 and its position in the colour-magnitude diagram further asserts that it does not belong to the cluster. However, an independent determination of the cluster membership for these two W UMa binaries is important in establishing their nature with respect to the cluster NGC~6866.

 \begin{table}
    \caption{The Photometric parameters obtained for the two W UMa binaries ID487 and ID494 using the WD method.}
    \begin{center}
\begin{tabular}{lccccc}
\hline
Elements                        &&     ID487               &&    ID494                   \\
\hline                                                      
\\                                                          
$T_1$ (K)                          &&     5160 	          &&   4805	             \\
$T_2$ (K)			        &&     5173$\pm$43	  &&   5026$\pm$37            \\
Spectral Type			&&      K0		  &&   K3                     \\ 
$q$		                &&    1.29$\pm$0.04       &&    0.39$\pm$0.02          \\
$i_0$		                &&   63.5$\pm$0.6         &&    68.4$\pm$0.7           \\
$\Omega_{1,2}$			        &&   4.103$\pm$0.056      &&    2.56$\pm$0.05          \\
$f$			 	&&   0.21	          &&   0.48                   \\
$r_1$        Pole 		&&   0.347$\pm$0.002      &&    0.452$\pm$0.010        \\
             Side 		&&   0.366$\pm$0.002      &&    0.488$\pm$0.014        \\
             Back 		&&   0.405$\pm$0.004      &&    0.521$\pm$0.020        \\
$r_2$        Pole		&&   0.390$\pm$0.002      &&    0.297$\pm$0.015        \\
             Side		&&   0.413$\pm$0.003      &&    0.312$\pm$0.019        \\
             Back		&&   0.449$\pm$0.004      &&    0.360$\pm$0.038        \\
$L_1$				&&   0.440	          &&   0.640                  \\
$L_2$				&&   0.560	          &&   0.360                  \\
$L_3$				&&   0.0		  &&    0.0                    \\
$x_1$				&&   0.60 	          &&   0.60                   \\
$x_2$				&&   0.60 	          &&   0.60                   \\
$A_1$				&&   0.50 	          &&   0.50                   \\
$A_2$				&&   0.50 	          &&   0.50                   \\
$G_1$				&&   0.32 	          &&   0.32                   \\
$G_2$				&&   0.32 	          &&   0.32                   \\
\hline
\end{tabular}
\end{center}
    \label{Tab2}
\end{table}

\section{Discussions}
%
\begin{figure*}
\centering
\vspace{0.5cm}
\includegraphics[width=6.5cm,height=4.7cm,angle=0]{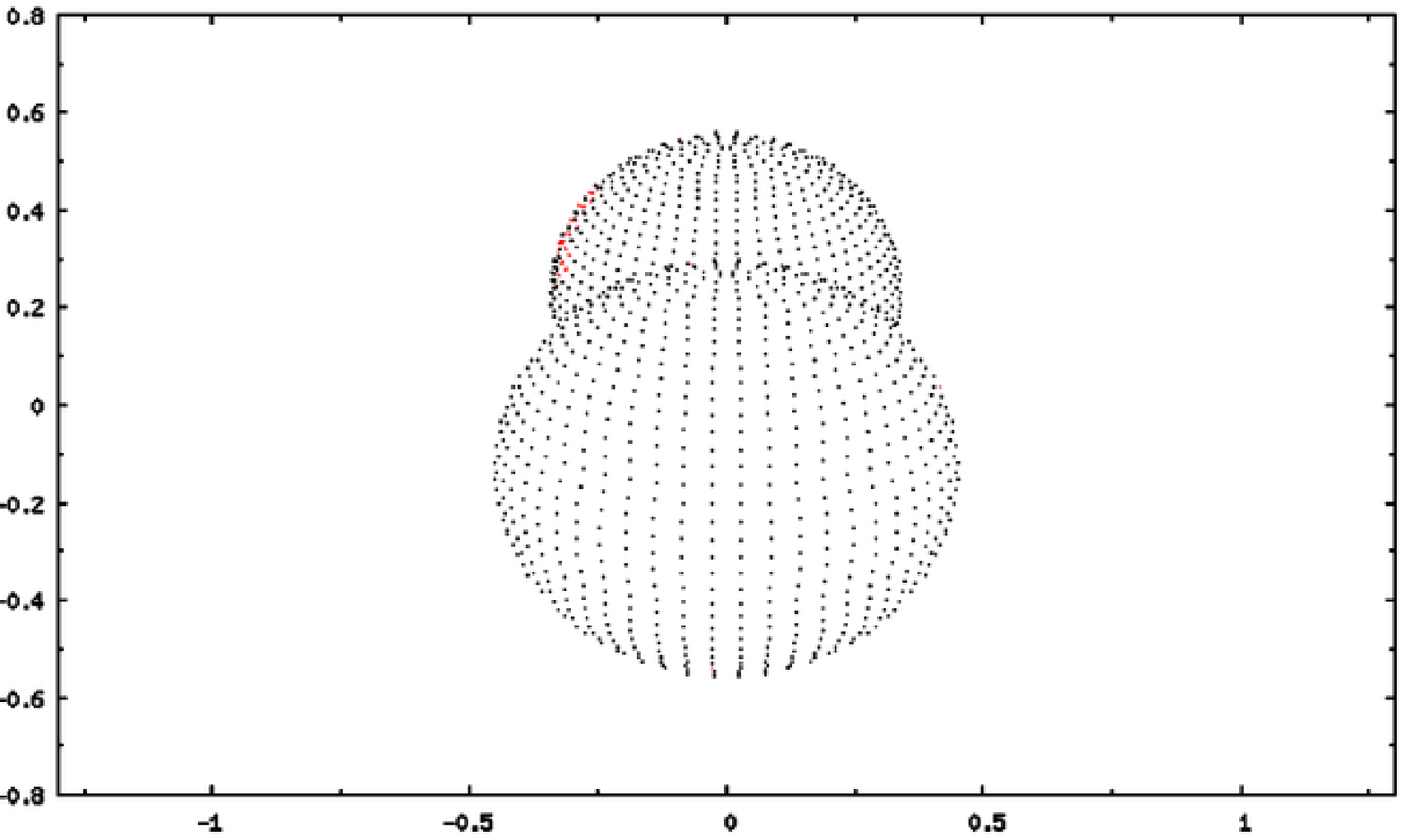}
\hspace{0.5cm}\includegraphics[width=6.5cm,height=4.7cm,angle=0]{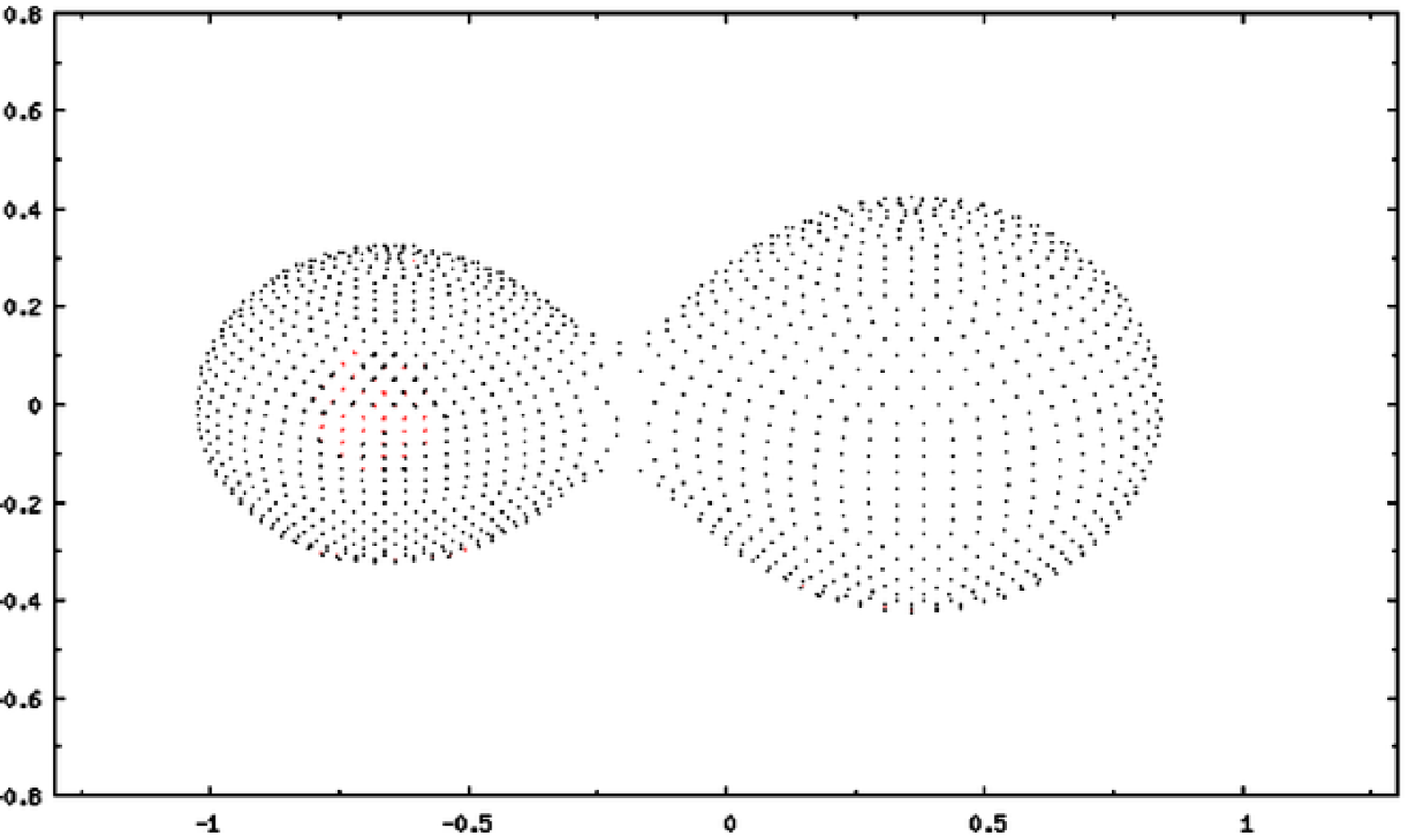}
\includegraphics[width=6.5cm,height=4.7cm,angle=0]{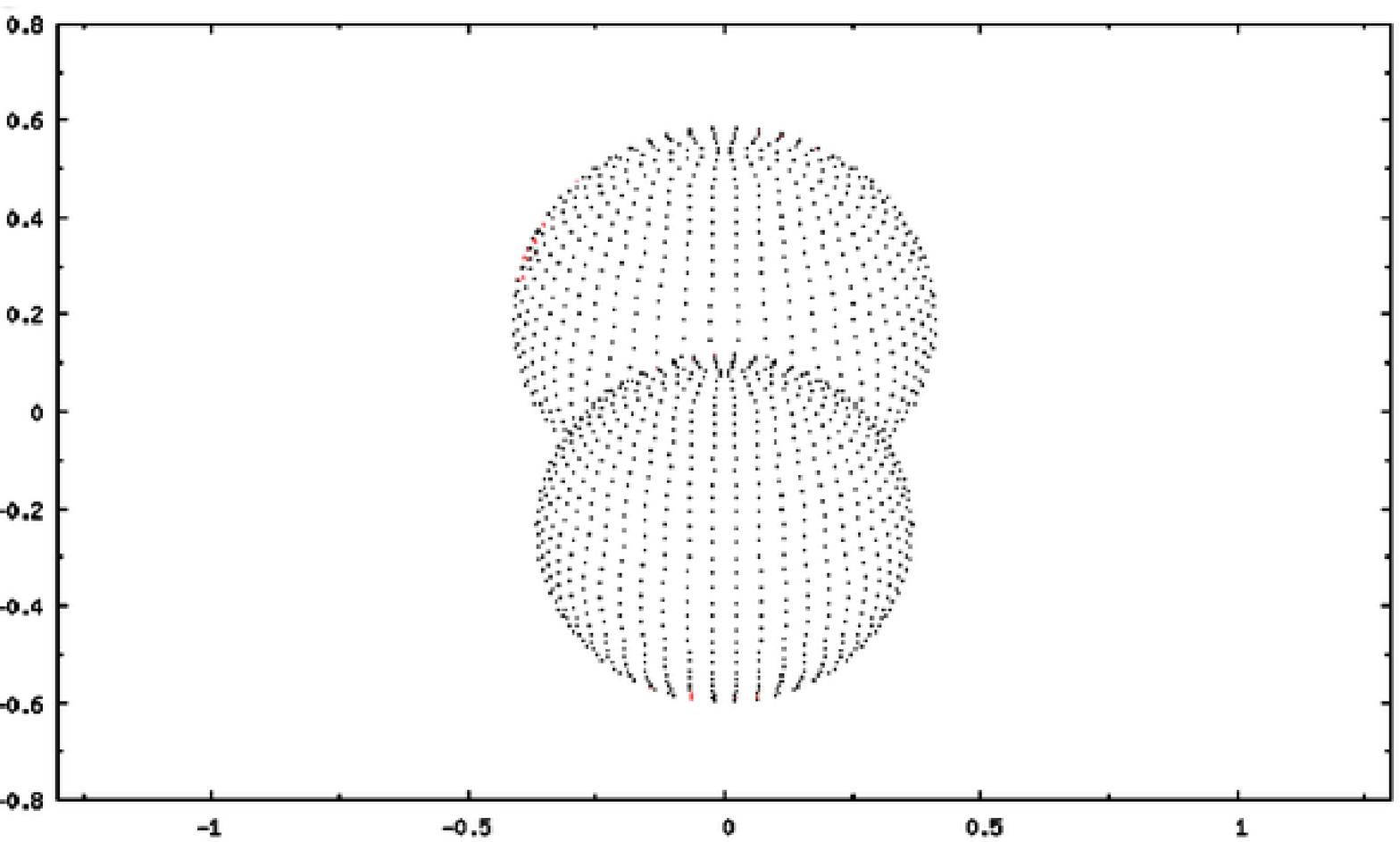}
\hspace{0.5cm}\includegraphics[width=6.5cm,height=4.7cm,angle=0]{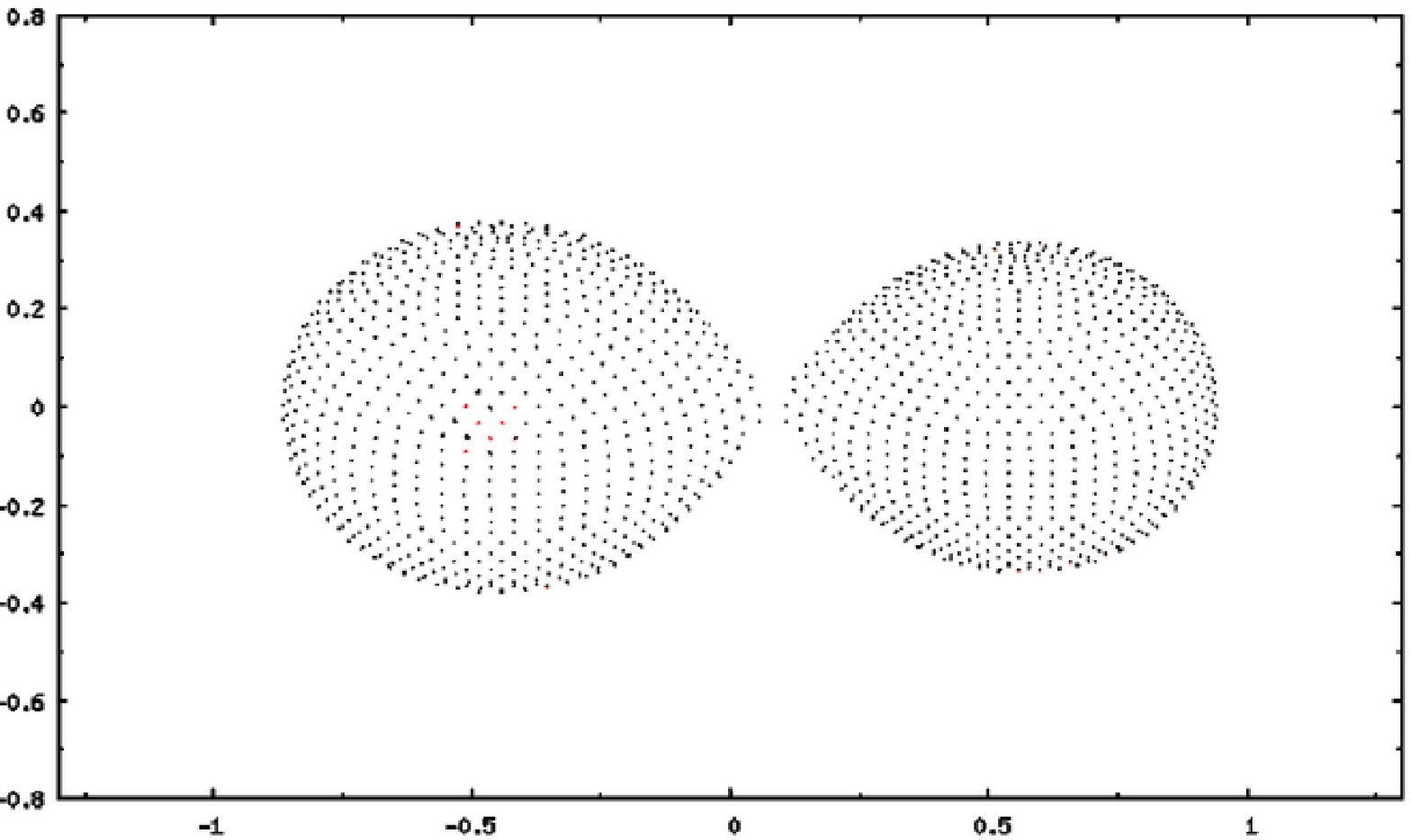}
\begin{flushleft}
\caption{Geometric configurations for the star ID487 (lower panel) and ID494 (upper panel) at phase 0.50 (left side) and 0.75 (right side).}
\label{Fig7}
\end{flushleft}
\end{figure*}

The light curve of binary ID487  shows similar maxima and minima except at the phase 0.75, in both the pass bands $V$ and $I$, which could be due to O'Connell effect. The light curve of binary ID494 also shows dissimilar maxima and minima suggesting a strong O'Connell effect. The spot parameters obtained from the solutions for both the binaries are shown in Table~3. Having a fill out factor of 21\% for the star ID487 and 48\% for the star ID494, and a small temperature difference between the components of about 13 K and 221 K in these two stars, it can be concluded that there is good thermal contact between the two components in both the systems and they are stable and least probable to merge (e.g., Kahler, 2004). The angle of inclination was found to be $\sim 63^o.5$ for the binary ID487  and $\sim 68^o.4$ for the binary ID494, indicating that both are partially eclipsing binaries.

In Figure~\ref{Fig7}, we show the geometrical configurations of the stars ID487 and ID494 at phases 0.50 and 0.75. Here, we have also considered O'Connell effect in both the binary systems which, in general, is associated with the presence of cool spots on one of the components (Kalomeni et al., 2007). However, asymmetries around the phases 0.50 and 0.75 could also be due to presence of hot spots, faculae, gas streams or any other inhomogeneity on the surface of the components (Yakut \& Eggleton, 2005). In most of the binaries the parameters of the spot vary with the activity cycle (Kose et al., 2011; Kalomeni et al., 2007), however, for the chosen binaries, the spot parameters do not show any variation over the period of observation. Specifically the large spot on one of the components of the W UMa binary ID494  seems to be persistent for the period of observation.

The following empirical relations were derived by Gazeas (2009), from the studies of several contact binaries.

$$
log M_1 = 0.725~log~P - 0.076~log~q + 0.365
$$
$$
log M_2 = 0.725~log~P + 0.924~log~q + 0.365
$$
$$
log R_1 = 0.930~log~P - 0.141~log~q + 0.434
$$
$$
log R_2 = 0.930~log~P + 0.287~log~q + 0.434
$$
$$
log L_1 = 2.531~log~P - 0.512~log~q + 1.102
$$
$$
log L_2 = 2.531~log~P + 0.352~log~q + 1.102
$$

Substituting the values of mass ratio and period from the best fit solution into these relations, we derived the mass, radius and luminosity of the binary components which are given in Table~4. Here, it should be noted that $q$ in the above equations represents mass of the lighter component divided by the mass of the heavier component of the binary system so in the case of ID487, $q$ was taken as 1/1.29.

\begin{table}
\caption{Spot parameters for the two W UMa binaries derived through WD method.}
\begin{center}
\begin{tabular}{lccccc}
\hline
Parameters            &&     ID487      &&     ID494     \\
\hline
\\
Longitude(radians)	&&	   4.599	&&	     4.599   \\
Latitude(radians)	&&	   1.25 	&&	     1.25   \\
Radius(radians) 	&&	   0.18 	&&	     0.40   \\
Temperature factor	&&	   0.80 	&&	     0.55  \\
$\Sigma(O-C)^2$		&&	   0.0352	&&	     0.0416  \\ 
\hline
\end{tabular}
\end{center}
\label{Tab3}
\end{table}

\begin{table}
\caption{The fundamental parameters obtained for the two W UMa binaries using the WD method and relations given by Gazeas (2009). All values are given in the solar units. Here, subscript "1" is used for higher-mass component and "2" for the lower-mass component of the binary system.}
    \begin{center}
\begin{tabular}{lcccc}
\hline
Parameters&&     ID487        &&     ID494     \\
\hline                                  
\\                                      
$M_1$    &&  1.24 $\pm$0.02 &&  1.20 $\pm$0.06  \\
$M_2$    &&  0.96 $\pm$0.05 &&  0.47 $\pm$0.01  \\
$R_1$    &&  1.24 $\pm$0.01 &&  1.22 $\pm$0.02  \\
$R_2$    &&  1.11 $\pm$0.02 &&  0.81 $\pm$0.01  \\
$L_1$    &&  1.55 $\pm$0.03 &&  1.62 $\pm$0.14  \\
$L_2$    &&  1.25 $\pm$0.08 &&  0.72 $\pm$0.01  \\
\hline
\end{tabular}
\end{center}
\label{Tab4}
\end{table}

\section{Summary}
We have presented the detailed $V$ and $I$ bands time-series photometric study of two W UMa binaries in the field of the distant cluster NGC~6866 which is located at a distance of about 1.5 kpc from the Sun. Both the binary systems are very faint having a similar magnitude of about 17.2 mag. The orbital periods of the stars ID487 and ID494 were found to be 0.415110$\pm$0.000001 day and 0.366709$\pm$0.000004 day, respectively. The photometric and infrared colours suggest a respective spectral class of K0 and K3 for ID487 and ID494. The photometric light curves were analyzed using the WD-code which provided mass ratios between the two components of the binary systems. Using the relations given by Gazeas (2009), photometric solutions for the binary systems yield $M_1 = 1.24 \pm0.02 M_\odot$ and $M_2 =  0.96 \pm0.05 M_\odot$ for the star ID487, and $M_1 =  1.20\pm0.06$ $M_\odot$ and $M_2 =  0.47\pm0.01$  $M_\odot$ for the star ID494. The light curves of these binaries show uneven maxima which is the signature of the O'Connell effect in the binary systems. The disagreement between two maxima of the light curves could be explained by the presence of a spot on the convective envelope of the secondary components. These two binary systems show a contact degree of $f$ = 21\% and 48\% with a small temperature difference between the two components as $\Delta T$ = 13~$K$ and 221~$K$ respectively. Orbital period investigation plays a vital role in understanding the evolution and searching for any third component in the stellar system. Although these binary systems were observed over a period of 3 years in the present study but it is important to have additional long-term photometric and possibly spectroscopic observations of these binaries that can help in understanding the changes in spot parameters with the associated magnetic activity cycle, if any, present in the stars. This shall further help to investigate any kind of variation in their orbital periods and mass transfer rate between the two components of the systems.
\begin{acknowledgements}
We are thankful to J. Molenda-\.{Z}akowicz for providing photometric data of their published
variables. Thanks to Rama Kant, N. K. Chakradhari, Manoj Patel and Vindor Kumar for their
observational support. YCJ acknowledge the financial support from the project DST/INT/SA/P-02 and RJ acknowledge the financial support from the project UGC-BSR research Start-Up Grant Sanctioned vide UGC Order No. F. 30-108/2015(BSR) of UGC, under which part of the work has been carried out.
\end{acknowledgements}


\label{lastpage}

\end{document}